# A numerical study on the design trade-offs of a thin-film thermoelectric generator for large-area applications

Kirsi Tappura*

VTT Technical Research Centre of Finland, P.O. Box 1300, FI-33101 Tampere, Finland

*to whom correspondence should be addressed

E-mail address of the corresponding author:
kirsi.tappura@vtt.fi

**Abstract**

Thin-film thermoelectric generators with a novel folding scheme are proposed for large-area, low energy-density applications. Both the electrical current and heat transfer are in the plane of the thermoelectric thin-film, yet the heat transfer is across the plane of the module – similar to conventional bulk thermoelectric modules. With such designs, the heat leakage through the module itself can be minimized and the available temperature gradient maximized. Different from the previously reported corrugated thermoelectric generators, the proposed folding scheme enables high packing densities without compromising the thermal contact area to the heat source and sink. The significance of various thermal transport, or leakage, mechanisms in relation to power production is demonstrated for different packing densities and thicknesses of the module under heat sink-limited conditions. It is shown that the power factor is more important than *ZT* for predicting the power output of such thin-film devices. As very thin thermoelectric films are employed with modest temperature gradients, high aspect-ratio elements are needed to meet the – usually ignored – requirements of practical applications for the current. With the design trade-offs considered, the proposed devices may enable the exploitation of thermoelectric energy harvesting in new – large-area – applications at reasonable cost.

KEYWORDS: thermoelectric generator; large-area TEG; thin-film, in-plane heat transfer; FEM simulations; computational TEG design

# 1. Introduction

There are plenty of thermal gradients available around us, which could, in principle, be exploited for sustainable energy scavenging by thermoelectric means. However, a major hurdle for a wider use of thermoelectric (TE) devices for energy harvesting is the low efficiency, which - with the conventional devices – tends to result in a high cost per converted power [1-2]. In addition, the commercially available thermoelectric devices are rigid and often relatively bulky, which limits their usage in many potential large-area applications apart from the cost. Further, most of the devices are intended for relatively large temperature gradients for which the heat leakage through the thermoelectric module itself is not as critical as for smaller gradients, while the latter are more readily available in our everyday environment.

Nevertheless, a significant number of various thermoelectric modules are commercially available, although the cost per converted power is not satisfactory for many applications. A popular approach to improve the cost competitiveness of the TE generators (TEG), is to concentrate on developing the properties of the TE materials to enhance the efficiency, or to increase the figure of merit, $ZT$ ($ZT = S^2\sigma T/k$, where $S$, $\sigma$, $T$ and $k$ are the Seebeck coefficient, electrical conductivity, absolute temperature, and thermal conductivity, respectively) [3]. As $ZT$ relates to the upper limit of the power conversion efficiency through the material properties, it is essential for the device optimization. However, it does not take into account the material consumption, fabrication costs, heat sink requirements or the fact that the device structure has an impact on the thermal transport mechanisms and paths [4]. Thin-film TE materials are attractive due to their potential compatibility with low-cost fabrication methods, such as roll-to-roll processing or screen printing, and lower material consumption. Flexible TEG designs, based on both organic [5-6] and inorganic [7, 8, 9] thin and thick TE films, have been studied for their potential application with arbitrary shape heat sources. The heat flux in the common thin-film TE modules is, however, perpendicular to the plane of the film, which significantly limits the temperature gradient available for power production due to the considerable thermal transport through the thin film, unless very efficient heat sinks are used [10].

Typically, TE modules consist of several n- and p-type bulk material or thin-film legs assembled as a circuit with the heat flux (temperature gradient) perpendicular to the plane of the TE module and the TE films [1, 7, 11-12]. For the time being, there are no commercially available TE modules where the heat flux occurs in the plane of a thin TE film, even though such an architecture allows significantly larger temperature gradients. However, there are several publications of various designs of the thin-film thermoelectric modules with the heat transfer in the plane of the TE film utilizing planar substrates [13, 14, 15, 16, 17, 18, 19], corrugated structures [20, 21, 22, 23, 24, 25] or different solid supporting architectures [26, 27] with some interesting approaches for flexibility and cost reduction [28, 29, 30, 31]. They all, except some of the organic or hybrid devices [13, 25, 31], are designed to consist of two different TE materials to form the basic unit (TE couple) of the thermoelectric module. This traditional design employing complementary n- and p-type materials allows convenient geometries for connecting the TE legs electrically in series and thermally in parallel, but forces one to use both the n- and p- type materials in spite of their quality, which may be a challenge especially for organics devices.

Many of the reports dealing with the thin-film TE generators with in-plane heat transfer, demonstrate the device performance by applying powerful heat sinks [26, 27] or forced temperature gradients [13, 22, 28-30] without considering the true effect of the various parasitic heat transfer mechanisms occurring in the device. However, the parasitic effects may have a significant impact on the temperature gradient, and thus, on the available power under heat sink-limited conditions. Although there are also more detailed thermal analyses performed for different planar in-plane thin-film TE structures [14-19], they do not provide insight into the designs employing a folded structure. In ref. [21] the authors present a well-defined theoretical performance optimization for a thin-film TE generator with a corrugated architecture under heat sink-limited conditions, but neglect the influence of heat conduction through the air cavity of the device. In the present paper, it will be shown that such parasitic effects, often ignored, may have a significant impact on the temperature gradient and, thus, on the power production under natural convection, or heat sink-limited conditions, in the thin-film devices with heat transfer in the plane of the TE film fabricated on a thin substrate with low thermal conductivity. Further, as the available electrical current is usually not an issue in the bulk or thin-film TE generators where the current flows across the plane of the film, its specific influence on the device performance is normally not considered in the related reports. Unfortunately, this seems to be the case even when reporting on the thin-film TE devices with the current flow in the plane of the thin-film. The fact that the electrical resistance may become very high in the TE modules consisting of several such thin-film TE units connected electrically in series, is usually well recognized [5, 23, 32]. However, the significance of the available electrical current is typically not discussed in the reports, even though the produced current may be too small for practical applications, especially under modest temperature gradients [18, 19, 21-25, 27].

In this paper, a novel folding scheme is proposed for the thin-film thermoelectric generators with the heat flux and electrical current parallel to the film surface but the temperature gradient perpendicular to the plane of the TE module. Different from the previously reported corrugated thermoelectric generators [21-25], the proposed folding enables a high packing density for the thermoelectric elements without compromising the thermal contact area to the heat source and sink. It also enables the application of only single conduction-type semiconductors as the TE materials and, thus, provides a possibility to simplify the manufacturing process and to avoid the usage of the lower quality TE materials in the module. The work concentrates on very thin TE films (400 nm) and modest thermal gradients (10-20 K) with natural convection on the cold side of the TE module without additional heat sinks, resulting to heat sink-limited conditions. Special attention is paid to the heat leakage mechanisms of the modules of different geometries and packing densities and to the impact of the electrical resistance and, in particular, the produced electrical current, on the usefulness of the TE modules for practical applications. It is shown that it is not just the maximum output power and voltage or the resistance of the module that define the performance. Instead, the available current should be of special concern when designing in-plane thin-film TE generators for specific applications.

Further, the small thickness of the TE film has some implications for the conventional design parameters. For example, *ZT,* describing the material properties, is not always a good measure for predicting the power production properties of such thin-film devices.

The aim is to demonstrate the performance and design trade-offs of the novel thin-film TE modules that can be achieved at minimal cost and are suitable for large-area, low energy-density, applications, such as energy harvesting on (or in) walls and windows or other surfaces providing appropriate temperature gradients with its environment. Deep understanding of the complex dependencies of the various design parameters can only be deduced from careful numerical studies as presented in this paper. The temperature gradients selected for this study were deduced from the long-term temperature measurements performed over and on the window glasses of VTT premised in Espoo, Finland, during a two-year period. The proposed TE module with the novel folding geometry suits well for this kind of large-area applications.

## 2. Materials and methods

### 2.1 Device architecture

The proposed TE module consists of a thin polymer, or any insulating flexible, substrate on which the planar thin-film TEGs can be fabricated with low-cost methods. A number of planar thin-film TE elements are connected electrically in series on the substrate. After the appropriate folding of the substrate, the three-dimensional (3D) TE module is formed with the TE elements connected thermally in parallel. A proposed configuration is shown in Fig. 1.

The module before folding the sheet is shown in Fig. 1a with the TE elements (or legs or TEGs) connected electrically in series (the electrical conductor lines shown in grey). The green arrows indicate the direction of the flow of the charge carriers (electrons for n-type materials, as in this study). The basic principle of operation can be understood as follows (see e.g. refs. [25, 31] for previously demonstrated single conduction type TE modules): The faster hot charge carrier (here electrons) diffuse further than the cold ones, which results to a net build-up of electrons - and, thus, a negative potential - at the cold end of each TE leg, leaving a positive potential at the hot end. In order to allow the directional charge transport and adding up the voltages, the elements are connected electrically in series by connecting the cold end (here the negative potential) of each leg to the hot end (positive potential) of the next leg with the narrow conductor lines (the grey lines in Fig. 1a). Finally, when a load is connected across the cold (-) and hot (+) ends of the TE chain as shown in Fig. 1a, the voltage produced by the TE module will cause the current to flow through the load generating electrical power. The same basic idea to connect the elements can be followed to create a module with different numbers of rows and columns. Similar TEG patterns can be printed on both sides of the substrate in order to double the TEG density. A 3D sketch and the side view of the folded module (viewed from the left side of the sheet of Fig. 1a after folding) are shown in Fig. 1b and 1c, respectively. Fig. 1c also depicts the possible supporting, thermally conductive structures - serving also as the thermal interfaces to the heat source and heat sink. When the sheet of Fig. 1a is folded along the dashed lines following the folding directions indicated in the figure, a profile shown in the side view is formed and the TEGs are assembled thermally in parallel and electrically in series in the module.

Although the bends does not need to be as sharp as shown in the sketches of Fig. 1b-c, relatively tight bends are required to form a folded structure with high TEG density (bending radius < 1 mm preferred). Because TE materials may not be able to flex according to the required bending radii without fracturing [33], the bending lines are designed to follow the conductor lines and, thus, the bends can be modified to stay inside the conductor lines, if needed. For dense packing, $t_{gap}$ can be made very small (≤ 50 μm) - as assumed in this paper - by filling the gap with a thin layer of electrically insulating glue. In order to use the glue also for attaching the folded substrate to the supporting structures or to the surfaces of the heat source/sink, the glue should also have high thermal conductivity.

From the resistance point of view, the width of the conductor lines (Fig. 1a) can be adjusted relatively freely, as long as the resistance of the interconnections can be kept low enough by controlling the conductor thickness, and the contact area to the TE material large enough for a sufficiently low contact resistance. Wide lines between the heat source/sink and the TE elements (the horizontal grey lines in Fig. 1a and short grey lines in Fig. 1c) are preferred for the quality of the thermal contact and heat spreading. However, the vertical conductor lines shown in Fig. 1a connecting the hot and cold side of the elements, should be kept sufficiently narrow (and/or thin) to mitigate the effect of the parasitic thermal conduction through the lines. To optimize the thermal contact and minimize heat leakage, the thickness and material properties of the flexible substrate and possible supporting structures should also be considered. Films, with good thermal conductivity at the high and low temperature contacts, are clearly preferred. However, the thermal conductivity, as well as the thickness, of the substrate under the TE material should be as low as possible in order to minimize the thermal leakage through the substrate between the hot and cold surfaces. Depending on the application and the fabrication constraints, the optimal folding profile may vary from sharp to round and from sparser to very dense. The folding scheme shown in Fig. 1 b and c enables both high-density TE modules and good thermal contacts – with a fairly large and adjustable contact area – to the hot and cold surfaces. For example, a square-wave folding (i.e. $\alpha = 90°$) would decrease the density of TEGs almost to a half of that shown in Fig. 1c, if the thermal contact area were kept the same. For $\alpha > 90°$, approaching the designs of refs. [20-24], the density would decrease even more.

Unless differently specified, the thickness of the TE material ($d_{TEG}$) is 400 nm, the thickness of conductors ($d_c$) 4 μm, the width of the conductor lines connecting the hot and cold sides ($w_c$) 6 mm and the thickness of the folded polymer substrate ($d_{subs}$) 25 μm. The length ($L_{TEG}$) and the aspect ratio ($AR = W_{TEG}/L_{TEG}$) of the TE elements are varied between 3.05 and 8.26 mm and 10 and 20, respectively (see Fig. 1a). In the calculations, the TEGs are assumed to be fabricated only on one side of the folded substrate.

*2.2 Computational methods and parameters*

The computational models built are based on the combination of the finite element method (FEM) and analytical calculations with self-written software. Three-dimensional FEM models are built for each of the basic repeatable TE units of the selected geometries comprising a two-leg TE generator with the necessary interconnections and folded substrate. Any two adjacent TE legs on the same row in Fig. 1a can be considered as the basic TE unit (before folding). Analytical methods are used for calculating the selected characteristic parameters of the TE devices as well as for multiplying the FEM model to a TE module consisting of hundreds or thousands of the units modelled with FEM.

In the FEM model, heat transfer equations are coupled with the electrical phenomena for modelling the thermoelectric effect (Peltier-Seebeck-Thomson) [34]. The FEM model includes the following coupled phenomena: heat transfer by conduction in the folded substrate, air, thermoelectric materials and conductor lines; electrical conduction and Joule heating in the TE material and conductor lines; thermoelectric effect in the TE material and conductor lines; temperature gradients applied over the various folded structures by setting a constant heat source or constant temperature on the hot side and convective heat transfer between the cold side of the TE module and the ambient air or, if so specified, by setting a constant temperature on both sides. The influence of convection and radiation within the TE module is assumed to be insignificant due to the low temperature differences studied. In spite of their potential importance, all the electrical contact resistances are also assumed to be negligible in this study. A perfect thermal contact is assumed between the horizontal conductor lines of Fig. 1a and the heat source and sink.

To evaluate the electrical characteristics, the FEM modelled TEG unit is connected to an external load resistor ($R_{load}$), and the simulations run with the different $R_{load}$ values are used to produce the current ($I$) – voltage ($U_{out}$) curves (IV curves). From the IV curves the output power ($P_{out}$) as a function of current and the internal resistance of the two-leg TEG unit ($R_{TEG}$), can be extracted by means of simple curve fittings:

$$U_{out}(I) = aI + b \rightarrow a = slope;\ b = U_{out}(0) \quad (1)$$

$$P_{out} = U_{out}I = aI^2 + bI \quad (2)$$

From the equations it can be seen that $|a| = R_{TEG}$ and $b = V_{oc} = S_{eff}\Delta T$, where $R_{TEG}$ is the resistance, $V_{oc}$ the open circuit voltage and $S_{eff}$ the effective Seebeck coefficient of the whole TEG device included in the FEM model. $\Delta T$ is the temperature gradient over the TE legs available for power production.

According to the traditional analysis, a TE generator can be operated at maximum power when $R_{load} = R_{TEG}$, (load matched) [1]. However, more recently it has been pointed out that as the electrical current depends on $R_{load}$, the additional heat generated or transferred by Joule heating and the Peltier effect is also influenced by $R_{load}$ [10, 35]. This means that increasing $R_{load}$ tends to increase $\Delta T$ and, therefore, the maximum power condition is not satisfied exactly at $R_{load}/R_{TEG} = 1$ but somewhat above 1. As the present work concentrates on heat sink-limited systems, and not for constant $\Delta T$s, the dependence of $\Delta T$ on $R_{load}$ was studied by the FEM simulations where all the essential phenomena are included as explained above. It was found that for all the relevant values of $R_{load}$ the change in $\Delta T$ was always less than 0.5 %. This concludes that the effect is very small for the thin in-plain TE elements and that the traditional impedance matching condition ($R_{load} = R_{TEG}$) for the maximum power ($P_{max}$) is a good approximation for the studied systems:

$$P_{max} = \frac{b^2}{4a} \qquad (3)$$

This is also supported by the results presented in section 3.1.

The sensitivity of the results of the FEM model to the size of the grid was found small: the change in all the values of interest remained $\leq 1$ % when doubling the mesh density. Before starting the production runs, the reliability of the basic FEM model for the thin-film TEGs was validated against the experimental results of ref. [36]: the simulated results were in good agreement with the experimental ones (error < 5%) and further improved when a contact resistance was added between the TE material and the metal in the test model.

After extracting the characteristics of the FEM modelled TEG unit consisting of a repeatable two-leg system, the corresponding data for a multiunit TEG module is obtained by analytical calculations. For example, the maximum power for a TEG module consisting of n units coupled electrically in series and thermally in parallel is

$$P_{nmax} = \frac{n(V_{OC})^2}{4R_{TEG}} = \frac{(V_{nOC})^2}{4R_{nTEG}}, \qquad (4)$$

where $V_{nOC} = nV_{OC}$ is the open circuit voltage of the module consisting of *n* units and $R_{nTEG} = nR_{TEG}$ the resistance of the module. If *m* such TEG modules are further connected in parallel, the total resistance of the module can be reduced to $R_{nmTEG} = \frac{n}{m} R_{TEG}$. This, however, comes at the expense of an m-fold increase in the total area required for the new module and, therefore, all TEGs are connected electrically in series in the studied configurations.

The material parameters listed in table 1 are used in the simulations, unless differently specified. Constant material properties are assumed realistic for the limited temperature range studied (i.e. no temperature dependence in the material parameters taken into account).

Unless stated otherwise, the temperature on the hot side of the device is forced to stay at a constant value ($T_h$), while the temperature of the ambient air on the other side of the device (cold side) is kept at temperature $T_{ca} = 23$ °C and convective heat transfer is assumed between the cold side of the device and the air. The heat rejection is assumed to be limited by free convection with the convective heat transfer coefficient $h_c = 5$

Wm$^{-2}$K$^{-1}$ on the cold side, unless differently specified. This assumption of natural convection is made to be able to analyze the performance of the system without elaborated heat sinks and, thus, to obtain the results for a setup with minimized costs. The initial temperature gradient is defined as $\Delta T_{init} = |T_h - T_{ca}|$, while the temperature gradient over the TE elements available for power production ($\Delta T$) is somewhat lower due to the various heat transfer mechanisms in the TEG system (i.e. the thermal conductivity, or thermal leakage, of the TEG device itself and the limited efficiency of convection for heat rejection).

The actual values of $\Delta T_{init}$ selected for the study were deduced from the temperature measurements performed over and on the window glasses of the premises of VTT in Espoo, Finland, during a two-year period. According to the measurements, the highest temperature gradients exist between the insulating double pane glasses – favoring the geometry of the proposed folded TEG, and the regularly occurring values of $\Delta T_{init}$ are in the range of 10-20 K or below with the polarity of the gradient changing frequently.

## 3. Results and discussion

*3.1 Current and power production and the effect of air convection*

From the application point of view, the first requirement for a TE generator is that a harvester electronics circuitry is able to operate with the power provided by the TE module. To be able to transfer the maximum power from the module to the electronics, the internal resistance of the TE module ($R_{nTEG}$) should be matched to the input impedance of the harvester electronics ($R_{load}$), as assumed in equation (4). However, it is not enough to get a certain output power from the TE module ($P_{nmax}$), as harvester electronics requires a minimum current to be obtained in addition to a minimum voltage. In the case of thin-film TEGs with the current running parallel to the film, the internal resistance of a TE element may become significant, and, thus, the produced current may not meet the requirements of the electronics (e.g. 10-80 µA depending the harvester circuitry [37-38] and the possible additional electronics needed e.g. due to the changing polarity of $\Delta T_{init}$). If the thickness of the TE material cannot be increased e.g. due to the fabrication constraints, TE legs with large aspect ratios ($AR = W_{TEG}/L_{TEG}$) are needed. In this study, where $d_{TEG}$ = 400 nm, $AR$ = 10 or more is needed to keep the internal resistance of a two-leg TEG unit below 14 Ω and the produced current above 80 µA for $\Delta T \geq 5$ K with the material parameters of Table 1. The resistance of the conductor lines is not an issue as far as the wirings can be made as thick as needed, which is usually the case. However, the influence of the contact resistance between the TE and conductor material – omitted in this study – may need special attention in practical realizations depending on the materials used.

In Fig. 2, the output power and voltage is depicted as a function of current and the $R_{load}/R_{TEG}$ ratio for a two-leg TEG unit with $AR$ = 15, as well as for a TE module consisting of the two-leg TEG units connected electrically in series and thermally in parallel according to the proposed design to cover an area of a square meter. The different curves correspond to the different convective heat transfer coefficients on the cold side and, thus, to the different $\Delta T$ available for power production, while $\Delta T_{init}$ is kept constant at 15 K. It can be seen that higher $\Delta T$ is preferred, not only for the higher power and voltage, but also for the higher current, while it is obvious that tiling several TE units in series only increases the power and voltage but not the current (compare Fig. 2a and Fig. 2c).

For a constant, and limited, temperature gradient and fixed material parameters, the required voltage and power can only be reached by connecting a number (*n*) of TEG units in series. Even if the produced current were sufficient, it is well known that the resistance of the module may become an issue, as it increases linearly with $n$ ($R_{nTEG} = nR_{TEG}$). As shown in Fig. 2d, the transferred power decreases rapidly, when $R_{nTEG}$ is considerably higher than the input impedance of the harvester electronics ($R_{load}$), while the impedance matching is obtained at $R_{load}/R_{nTEG} = 1$. For example, $R_{nTEG} \approx 31$ k$\Omega$ for the one-square-meter TE module of Figs. 2c-d, while the input impedance of the harvester electronics circuitries designed for converting power from high impedance DC sources may be of the order of 10 k$\Omega$ [37-38]. With these values, a power transfer efficiency of about 73 % is reached. If the size of the module, i.e. the number of TEG units in series, is doubled, the efficiency drops to 46 %, and for a four-square-meter TE module the efficiency is only 26 %.

To summarize, high $R_{TEG}$ can seriously limit the useful power generated from a specific temperature gradient, even if $P_{nmax}$ and the voltage at $P_{nmax}$ ($U_{@Pnmax}$) appear sufficient after connecting a number of TE elements electrically in series. It was shown that this relates to two facts: first, if the required minimum current is not reached, the harvester electronics cannot operate, although the produced power and voltage appear satisfactory. Second, even if the minimum current is reached, at some point when connecting more and more TE elements in series to produce more power (and a higher voltage), the total resistance of the TE module becomes too large for a sufficient power transfer to the electronics due to the poor impedance matching. As shown, these points may be of concern with the thin-film TEGs having the current flow in the plain of the film even with the present harvester electronics circuitries designed for high impedance DC sources.

*3.2 Influence of the aspect ratio and ZT*

In Fig. 3, the influence of *AR* and *ZT* on the power and voltage produced per a two-leg TEG and per unit area for a TE module consisting of the two-leg TEG units is shown as a function of current. Both the Seebeck coefficient and *AR* have a significant influence, not only on the output power and voltage, but also on the current (Fig. 3a). As expected from eq. (1), the increase in current is in the same proportion as the increase in *AR* (or decrease in $R_{TEG}$) for a constant *ZT*. However, the influence of the thermal conductivity of the TE material ($k_{TEG}$) is almost negligible: When $k_{TEG}$ is decreased from 3.5 W/m/K to 1.5 W/m/K and even to 0.57 W/m/K, only very minor increase in the power and IV curves can be seen, in spite of the fact that *ZT* changes, respectively, from 0.21 to 0.50 and finally to 1.32 for $S = -250$ μV/K. Similarly for $S = -90$ μV/K, the change of $k_{TEG}$ from 3.5 to 0.57 W/m/K increases $ZT$ from 0.03 to 0.17, but the change in the power production and IV curves is hardly distinguishable. The minor influence of $k_{TEG}$ on the device properties can be explained by the very small thickness of the TE material, which as such limits the (in-plane) thermal conduction in it. Thus, instead of the TE thin-film, the substrate and other media of the TE module are mainly responsible for the heat leakage, or the $\Delta T$ available for power production, as will be further discussed in the next sections. It is also worth noting that the results presented above are in accordance with the conclusions of Alvarez-Quintana [39] and Yamamoto et al. [15].

In spite of the clear influence of $AR$ on the power production in Fig. 3a, there is no remarkable difference in the maximum power per unit area for $AR = 10$ and $AR = 15$ with the same $ZT$ in Fig. 3b. This relates to the fact that with higher $AR$ less TEG units are needed for the same power. So, lower $R_{TEG}$ (higher $AR$) and materials with a higher power factor ($PF = \sigma S^2$) – in addition to higher $\Delta T$ – are highly preferred. This is not only due to the higher output power and voltage, but also due to the higher current generated per a TEG unit, and, thus, due to the smaller number of TEG units needed for a module, which leads to smaller $R_{nTEG}$ as well.

The slightly different values of $\Delta T$ (~ 6.7 K and 6.8 K) for the TEGs with $AR = 10$ and $AR = 15$ can be explained by the different relative widths of the conductor lines and TE material between the hot and cold surfaces: $w_c/W_{TEG}$ = 1 mm/(10 × 5.05 mm) = 0.020 for $AR = 10$ and 1 mm/(15 × 5.05 mm) = 0.013 for $AR = 15$. The heat leakage through the conductor lines is, therefore, slightly more significant for the designs with $AR = 10$. On the other hand, even doubling $d_{TEG}$ has only a minor influence on the heat leakage.

In principle, the power generation of a TE module should improve if TE materials with higher $ZT$ are used. However, when considering very thin TE films as in the analysis above, the improvement may not be as significant as assumed from the difference in $ZT$s alone. For example, high performance $Bi_2Te_3$ [21] with $S$ = 236 μV/K, $\sigma_{TEG} = 8.13 \times 10^4$ S/m and $k_{TEG} = 0.57$ W/m/K has $ZT \approx 2.38$ at room temperature. If TEGs with $d_{TEG} = 400$ nm, $L_{TEG} = 5.05$ mm, $d_{subs} = 25$ μm and $AR = 15$ (similar to the dash-double dotted lines in Fig. 3b) were fabricated from $Bi_2Te_3$, $P_{nmax}$ per unit area would only be about 70 % larger than that of the dash-double dotted line in Fig. 3b according to simulations. This is the case even though $ZT$ for $Bi_2Te_3$ is more than 11 (2.38/0.21=11.3) times greater. Instead, the corresponding ratio of the power factors, i.e. 1.8, is in reasonably good accordance with the respective $P_{nmax}$ values.

*3.3 Comparison with a TEG composed of both the n- and p-type TE materials*

Fig. 4a compares the performance of the two-leg TE generators consisting of only one TE material (Fig. 4b) with the traditional TEG design (Fig. 4c) having one of the legs composed of n-type and the other of p-type

TE material [2]. Different electrical interconnections are needed for the two designs in order to connect the legs electrically in series (see Figs 4b and 4c). Otherwise, identical geometries similar to the ones in Fig. 3a are used for both the designs. It can be seen that both the power and current production is higher for the traditional (n/p) TEG designs (thick lines without markers in Fig. 4a). This is due to the somewhat higher $\Delta T$ as compared with that of the TEGs composed of only one type (n) TE material (thin lines without markers, except the dotted one). This relates to the fact that there are no conductor lines from the hot to the cold side in the traditional TEGs that causes the additional path for heat leakage in the single conduction-type TEGs. It should be emphasized, however, that these results are obtained by applying similar material properties (similar absolute values) for both the n- and p-type TE materials. In practice, one of the material types is often of lower quality, which may compensate or even invert the power production properties of the two TEG designs (see the dotted line in Fig. 4a as an example). The curves of fixed ΔT (7 K) show very similar power production and IV curves for both the designs (lines with markers).

*3.4 Influence of the packing density and the thickness of the substrate*

The maximum output power per unit area and the corresponding $\Delta T$ as a function of angle $\alpha$ are depicted for various TE module geometries under different conditions in Fig. 5. The thickness of the polymer substrate ($d_{subs}$) is either 25 or 130 µm, the length of a TEG unit ($L_{TEG}$) varies from 3.05 to 8.26 mm, $AR$ = 10 or 15 and $\Delta T_{init}$ is varied from 10 K to 20 K for the different designs (see the figure captions and legends for details).

A general trend in Fig. 5a-c is that the maximum output power per unit area increases as a function of angle $\alpha$, which can be explained by the increased density of the TEG units in the module – and by an additional impact from the simultaneous increase in $\Delta T$ for $\alpha < 60...70°$. In ref. [21] similar behavior for the maximum power was observed as a function of the corrugation angle. However, it was explained to be only a result of the increase in the thermoelectric element packing density, as their model ignored the thermal conductivity of the air cavity and did not provide information of $\Delta T$. The fact that ΔT slightly increases in Figs. 5 as a function of $\alpha$ seems obvious as the distance between the hot and cold surfaces increases when the thickness of the module [$t_{mod} = L_{TEG}\sin(\alpha)$] – and the thickness of air in the module – increases. However, the increase in the TEG density as a function of $\alpha$ also means that the relative density of the substrate material grows in the space between the hot and cold surfaces. This leads to increased thermal leakage over the module due to the high thermal conductivity of the substrate ($k_{subs}$ = 0.12 W/m/K) compared to that of air ($k_{air}$ = 0.0257 W/m/K). It can be seen that for $\alpha > 60...70°$ the latter effect starts to dominate in $\Delta T$, and $\Delta T$ starts to decrease, but $P_{nmax}$ per area continues its increase due to the continuing growth in the TEG density. Their relative behavior can be understood as follows: As $\alpha > 60...70°$ and approaching 90°, the growth rate of $t_{mod}$ approaches zero ($dt_{mod}/d\alpha = d[L_{TEG}*\sin(\alpha)]/d\alpha = L_{TEG}*\cos(\alpha)$), while the growth rate of the TEG density (~ $\sin(\alpha)/\cos^2(\alpha)$) is approaching its highest value.

The influence of the thermal conductance of the substrate can be seen in Figs. 5b and 5c, where the results are shown for TE modules simulated with the two different thicknesses of the substrate (25 and 130 µm). With the thinner substrate ($d_{subs}$ = 25 µm) the drop in $\Delta T$ at $\alpha$ > 60…70° is significantly smaller or not observable, while $P_{nmax}$ per area experiences the most significant increase at these angles when compared with the curves of $d_{subs}$ = 130 µm. This relates to the angle dependent growth rate of the TEG density (~ $\sin(\alpha)/\cos^2(\alpha)$) and the fact that the thinner substrate enables a slightly higher density of TEGs in the module, in addition to the lower density of the substrate material available for the heat leakage. The observed influence of $\Delta T_{init}$ is clear - and expected - on both the $P_{nmax}$ per area and $\Delta T$ giving higher values for higher $\Delta T_{init}$ (Figs. 5a and 5b).

To further investigate the influence of heat leakage through the air cavity of the folded module, the thermal conductivity of "air" was reduced to $1\times10^{-5}$ W/m/K (see the large squares in Fig. 5b at $\alpha$ =70°). It can be seen that the effect is significant, which suggests that the performance of the TE module can be improved by encapsulating the module in vacuum or filling the space with a gas of lower thermal conductivity.

Another interesting result can be seen by comparing Fig. 5b, where $L_{TEG}$ ~ 5 mm, and Fig. 5c, where $L_{TEG}$ ~ 3 mm. In spite of the lower $\Delta T$s – and, thus, the lower output power per a TEG unit - the output power per unit area is higher in Fig. 5c than in Fig. 5b for the same $\Delta T_{init}$, $\alpha$ and $d_{subs}$. This is simply due to the fact that the TEG density for the same $\alpha$ and $t_{gap}$ is higher for $L_{TEG}$ ~ 3 mm than for $L_{TEG}$ ~ 5 mm, as $L_c = L_{TEG} \cos(\alpha)$.

## 4. Conclusions

A new folding scheme for the thin-film thermoelectric generators with the heat flux and current flow parallel to film surface but the temperature gradient perpendicular to the plane of the TE module was proposed. Various design aspects were considered and the performance of different geometries analyzed computationally. In addition to highlighting the advantages of the proposed structure, critical discussions on the design challenges were given with the results of the simulated devices as demonstrators.

The high resistance of the thin-film TEGs was shown to set the most significant constraints to the device design and performance, in addition to the power factor. With the typical material parameters and the modest temperature gradients studied, TE elements with high aspect ratios are needed to be able to produce the current required by harvester electronics. This, however, may set some limitations to the usage, as the highest temperature gradients are often available over relatively long distances and, thus, a TEG module assembled of long TEG units may need large areas due to the required aspect ratio. The produced power is very sensitive to the design parameters of the folded structure and to the properties of the materials and intermediate media, but not to the thermal conductivity of the very thin TE films. As shown, the dependencies are not always obvious and, thus, careful application-specific computational optimization is required before moving to practical implementations.

The proposed folding scheme for the thin-film TE modules – with the air (or gas or vacuum) filled inner parts –provides a promising design for large-area applications. One of the advantages is that they can be attached either on a single warm surface located in a cooler environment (or the other way around) or in between two surfaces at different temperatures without elaborated, costly, heat sinks. Due to the potential to produce low-cost TE modules for large areas, the proposed designs may enable the exploitation of thermoelectric energy harvesting in new applications, e.g. on/in walls, windows or other surfaces providing low quality, but sufficient, temperature gradients with its environment.

**Nomenclature**

| | |
|---|---|
| 3D | three-dimensional |
| AR | aspect ratio = $W_{TEG}/L_{TEG}$ |
| $d_c$ | thickness of conductors |
| $d_{subs}$ | thickness of the folded polymer substrate |
| $d_{TEG}$ | thickness of the TE material |
| $h_c$ | heat transfer coefficient on the cold side |
| $I$ | electrical current |
| $k$ | thermal conductivity |
| $k_{TEG}$ | thermal conductivity of the thermoelectric material |
| $L_{TEG}$ | length of the thermoelectric elements |
| $L_c$ | = $L_{TEG}\cos(\alpha)$, width of the conductor lines parallel to the hot and cold surface of the module providing also thermal contact to the heat source and sink |
| $n$ | number of the thermoelelectric units |
| PF | power factor |
| $P_{max}$ | maximum output power of a thermoelectric unit |
| $P_{nmax}$ | maximum output power of a module of n thermoelectric units |
| $P_{out}$ | output power |
| $R_{load}$ | load electrical resistance |
| $R_{TEG}$ | internal electric resistance of a thermoelectric unit |
| $R_{nTEG}$ | internal electric resistance of a module of n thermoelectric units |
| $S$ | Seebeck coefficient |
| $S_{eff}$ | effective Seebeck coefficient of the TEG device |
| $T$ | absolute temperature |
| TE | thermoelectric |
| TEG | thermoelectric generator |
| $t_{gap}$ | gap between the adjacent folds |
| $t_{mod}$ | = $L_{TEG}\sin(\alpha)$, thickness of the thermoelectric module |
| $T_h$ | absolute temperature on the hot side |
| $T_{ca}$ | absolute temperature of the cold side ambient environment |
| $\Delta T_{init}$ | $\|T_h - T_{ca}\|$ |
| $\Delta T$ | temperature gradient over the TE legs available for power production |
| $U_{out}$ | output voltage |
| $U_{@Pnmax}$ | output voltage at the maximum power |
| $V_{oc}$ | open circuit voltage |
| $V_{nOC}$ | the open circuit voltage of a module consisting of $n$ thermoelectric units |
| $w_c$ | width of the conductor lines connecting the hot and cold sides |
| $W_{TEG}$ | width of the thermoelectric elements |
| $ZT$ | thermoelectric figure of merit |

*Greek symbols*

| | |
|---|---|
| $\alpha$ | stacking angle of the folded thermoelectric module |
| $\Delta$ | change in a property |
| $\sigma$ | electrical conductivity |
| $\sigma_{TEG}$ | electrical conductivity of the thermoelectric material |

**Acknowledgements**

This project has received funding from the European Union's Horizon 2020 research and innovation programme 2014-2018 under grant agreement No 645241. The partial funding from VTT Technical Research Centre of Finland Ltd is also gratefully acknowledged. K. Jaakkola and I. Marttila are gratefully acknowledged for making it possible to gain information of the available temperature gradients over and on the window glasses in Espoo, Finland.

**Figure captions**

Fig. 1. Design of the proposed TE module. (a) The TE module sheet before folding with a proposed way to assemble the TEGs on a single-sheet flexible substrate to enable the formation of a 3D TE module (red-blue gradients = temperature gradients in the TE materials, where red = hot, blue = cold; grey = electrical conductor lines, e.g. metal ink). $L_{TEG}$ indicates the length and $W_{TEG}$ the width of a TE leg, the dashed lines the approximate positions of the folds and words "up" and "down" in which direction the nearby sheet needs to be folded from the dashed line in question. The folds marked with letters A-L refer to the corresponding folds/edges in b and c after folding. The green arrows indicate the direction of the flow of the charge carriers (shown only for a part of the module for clarity). (b) The basic structure of the folded 3D thermoelectric module formed from the single planar sheet shown in a. (c) Side view of the folded structure ($W_{TEG}$ perpendicular to the plane of the paper). $t_{mod}$ indicates the thickness of the module and $L_c = L_{TEG}\cos(\alpha)$ corresponds to the width of the conductor lines (the horizontal conductor lines in a) and to the thermal contact to the heat source and sink. The free space between the hot and cold surfaces is assumed to be filled with air (light blue area in c). Note that the dimensions in the figures may not be in the optimal proportions.

Fig. 2. The output power (black curves) and voltage (red lines) as a function of (a, c) current and (b, d) the resistance ratio obtained from the FEM simulations for (a, b) a two-leg TEG unit and (c, d) for a TE module of the size of 1 m$^2$ consisting of 3456 couples of the two-leg TEG units (or of 6912 legs) connected in series according to the proposed design. $L_{TEG}$ = 5.05 mm, $\Delta T_{init}$ = 15 K, $AR$ = 15 and $\alpha$ = 70°. The solid, dashed,

dash-dotted and dash-double dotted curves correspond to the different $h_c$ (5, 10, 20 and 40 Wm$^{-2}$K$^{-1}$) on the cold side and, thus, to the different $\Delta T$ (~ 6.5, 9.3, 11.5 and 13.1 K, respectively) available for power production.

Fig. 3. The output power (black curves) and voltage (red lines) as a function of current (a) for a two-leg TEG and (b) per unit area for a module consisting of the two-leg TEG units with different $AR$s and $ZT$ varied by changing $S$ and $k_{TEG}$. $L_{TEG}$ = 5.05 mm, $d_c$ = 1 mm, $\Delta T_{init}$ = 15 K and $\alpha$ = 70°, $\Delta T$ ~ 6.7 K and 6.8 K for $AR$ = 10 and $AR$ = 15 K, respectively. $k_{TEG}$ = 3.5 W/m/K for the curves/lines without markers (A, B, C, D), $k_{TEG}$ = 1.5 W/m/K for the curves/lines with open markers (G, H) and $k_{TEG}$ = 0.57 W/m/K for the curves/lines with solid markers (E, F, I, J).

Fig. 4. (a) Comparison of the performance of the two-leg TEG units consisting of only one (n-type) TE material (n) with the traditional thermoelectric design employing both n-type and p-type TE materials (n/p). Schematics of the designs with the electrical interconnections for (b) the n-type and (c) the traditional TEG design. The green arrows indicate the direction of the electron flow. Otherwise, identical geometries similar to the ones in Fig. 3a are used for both the n and n/p designs. $k_{TEG}$ = 3.5 W/m/K for all the curves. $\Delta T_{init}$ = 15 K and natural convection with $h_c$ = 5 Wm$^{-2}$K$^{-1}$ is applied for the lines without markers. The lines with markers correspond to fixed $\Delta T$ = 7 K.

Fig. 5. Maximum output power ($P_{nmax}$) per unit area (solid black markers with black lines) and the corresponding final $\Delta T$ (open red markers with red lines) as a function of angle $\alpha$ (see Fig. 1) for the various TE module geometries under different $\Delta T_{init}$. (a) $d_{subs}$ = 130 µm, $L_{TEG}$ = 8.26 mm and $AR$ = 10. For $\Delta T_{init}$ see the legend. (b) $d_{subs}$ = 130 µm and $L_{TEG}$ = 5.26 mm for the triangles and circles, while $d_{subs}$ = 25 µm and $L_{TEG}$ = 5.05 mm for the squares and diamonds. $AR$ = 15. For $\Delta T_{init}$ see the legend. The large squares at $\alpha$ =70 ° correspond to a test with $k_{air}$ = 1×10$^{-5}$ W/m/K. (c) $\Delta T_{init}$ = 15 K. $d_{subs}$ = 130 µm and $L_{TEG}$ = 3.26 mm for the circles and $d_{subs}$ = 25 µm and 3.05 mm for the squares. $AR$ = 15. The lines are only drawn to guide the eye.

Table 1. Material parameters used in the simulations.

|  | TE material | Conductors | Polymer substrate | Air |
|---|---|---|---|---|
| Electrical conductivity, $\sigma$ [S/m] | $4\times10^4$ | $2.67\times10^7$ | | |
| Thermal conductivity, $k$ [Wm$^{-1}$K$^{-1}$] | 3.5 | 238 | 0.12 | 0.0257 |
| Seebeck coefficient, $S$ [μV/K] | -250 | 3.5 | | |
| Power factor, $PF = \sigma S^2$ [Wm$^{-1}$K$^{-2}$] | $2.5\times10^{-3}$ | $1.4\times10^{-6}$ | | |
| $ZT$ at room temperature | 0.21 | $4\times10^{-4}$ | | |

Fig 1 (a):

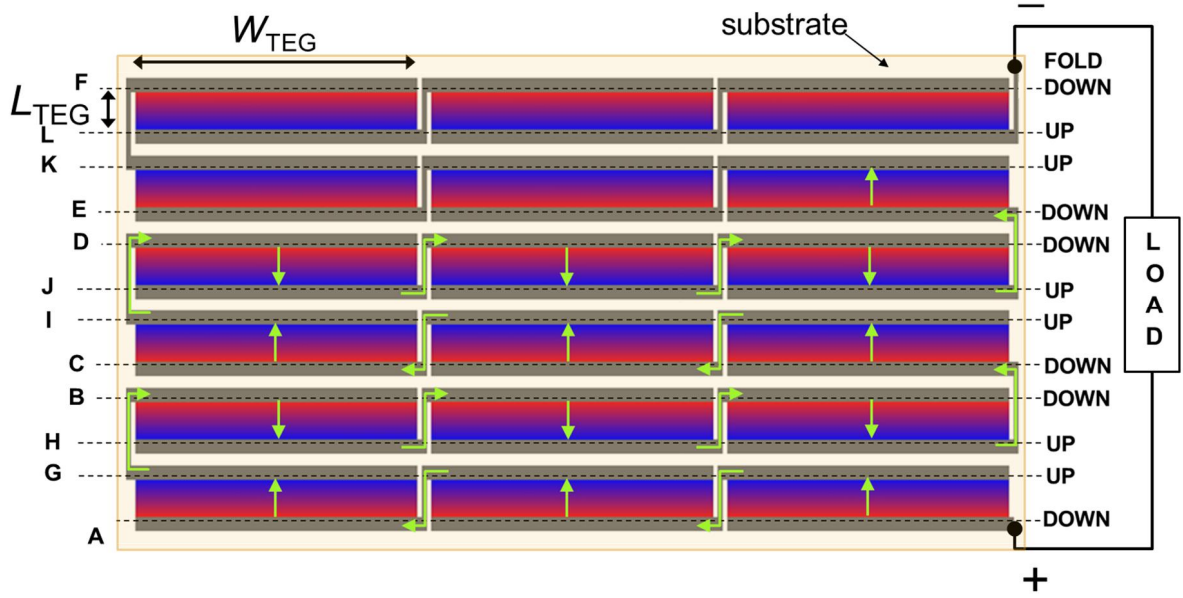

Fig. 1 (b):

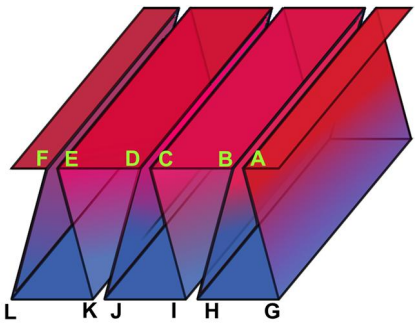

Fig. 1 (c):

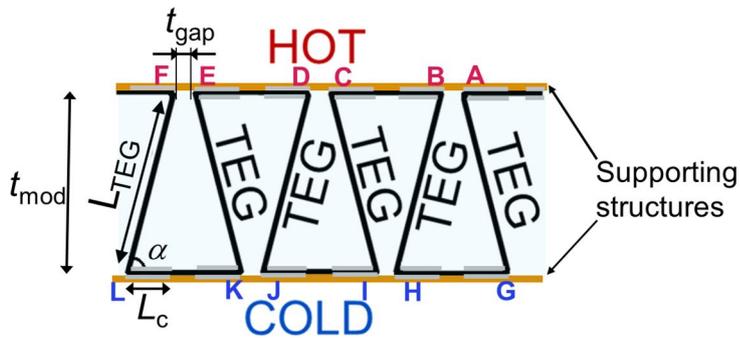

Fig. 2 (a):

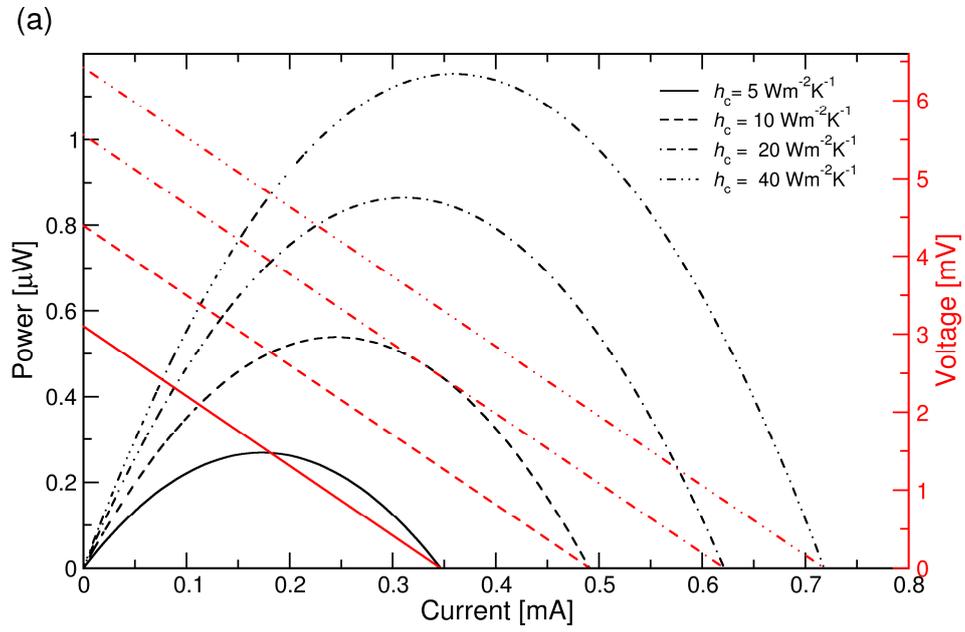

Fig. 2 (b):

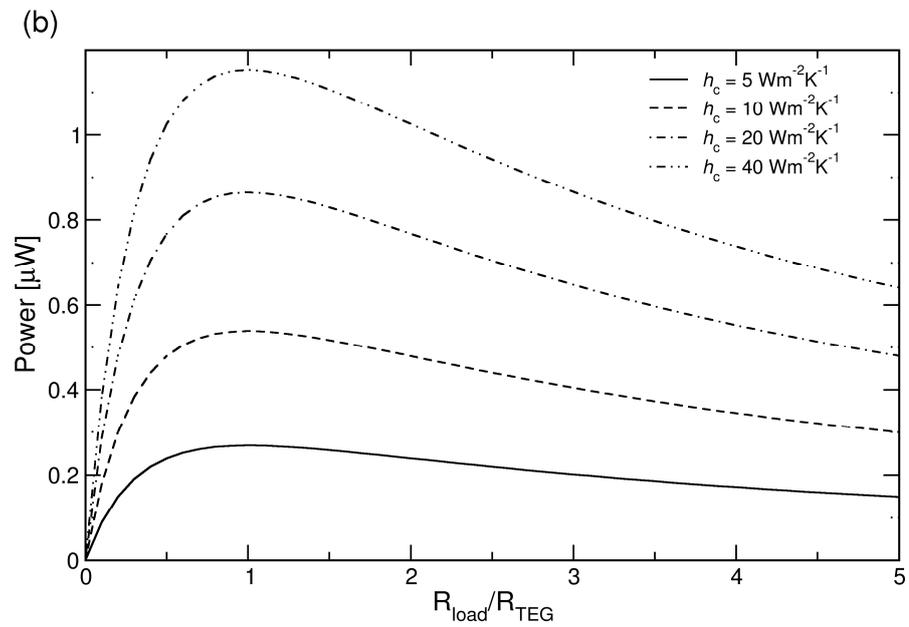

Fig. 2 (c):

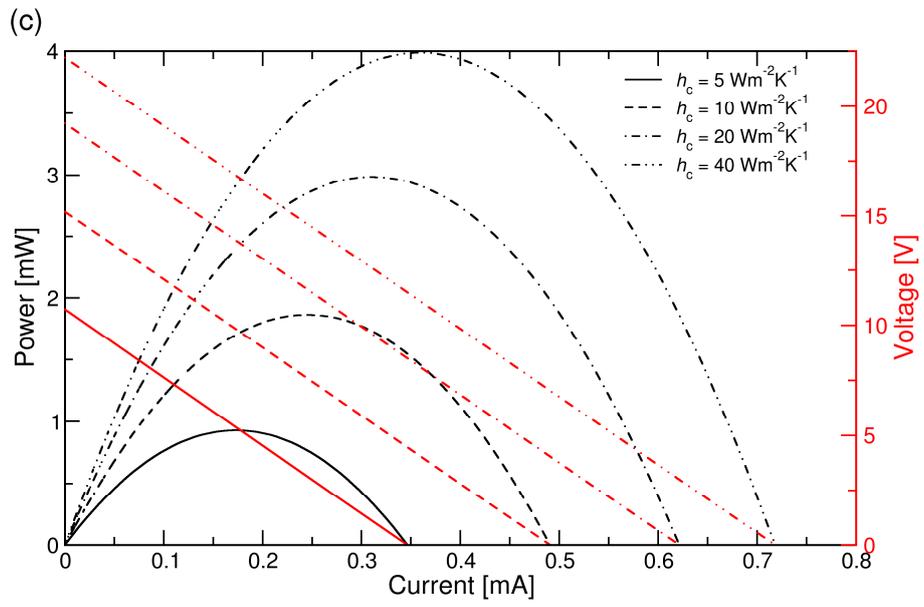

Fig. 2 (d)

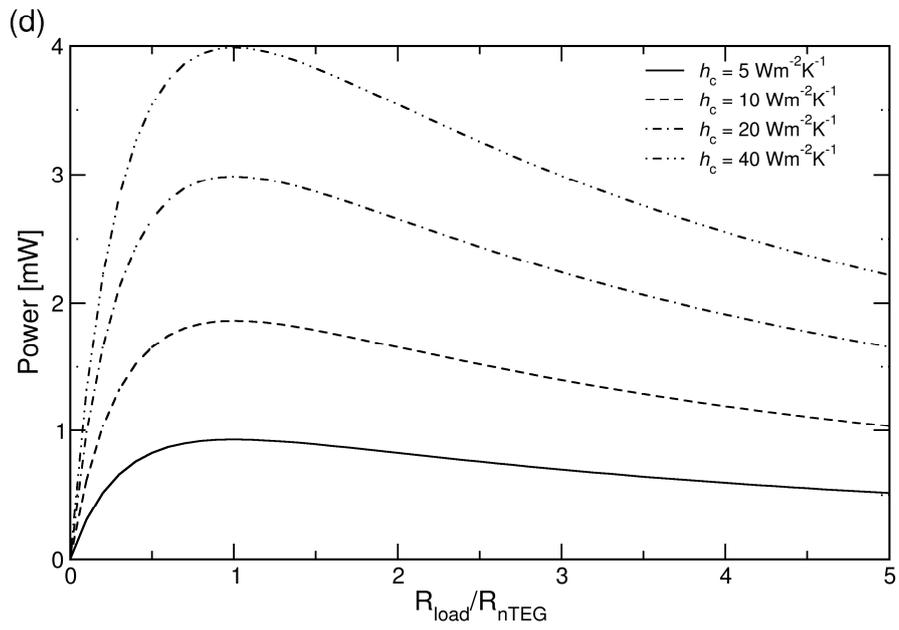

Fig. 3 (a):

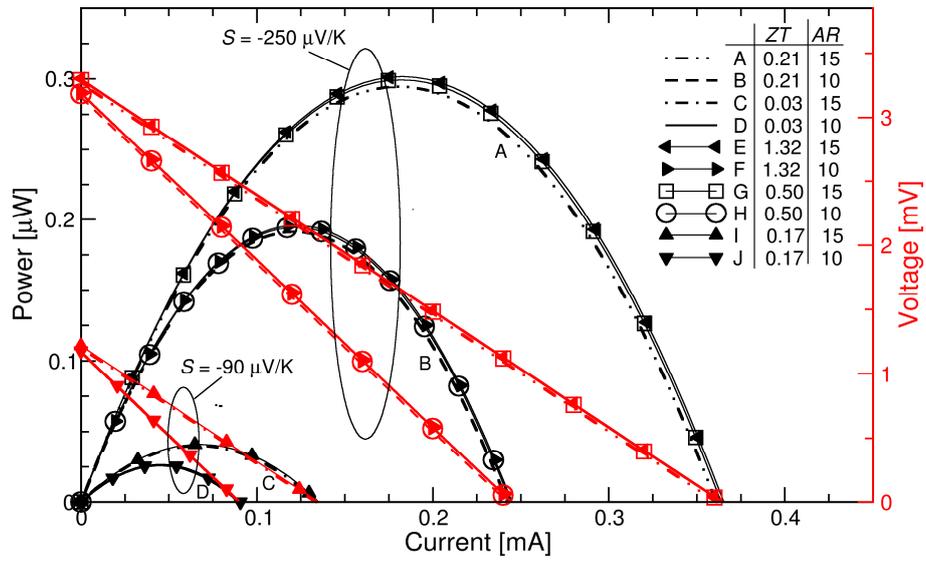

Fig 3. (b):

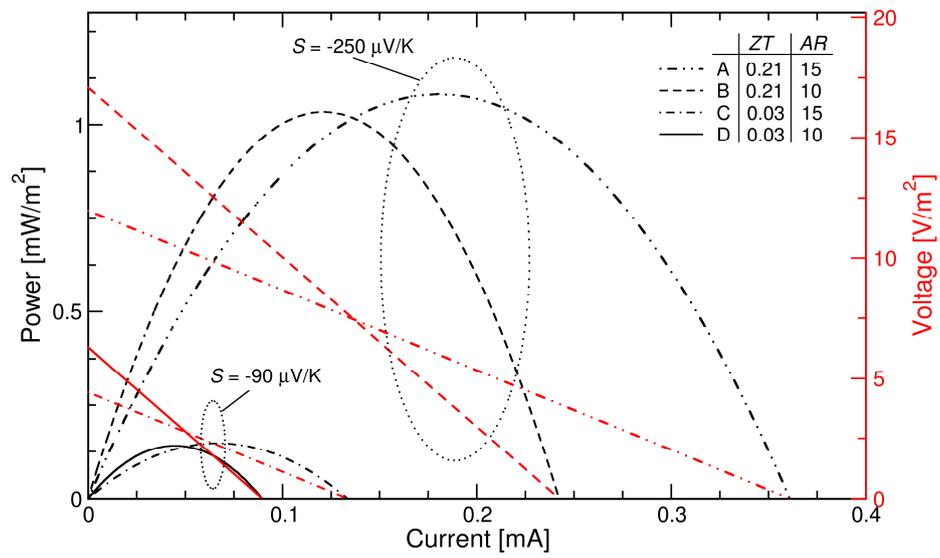

Fig. 4(a):

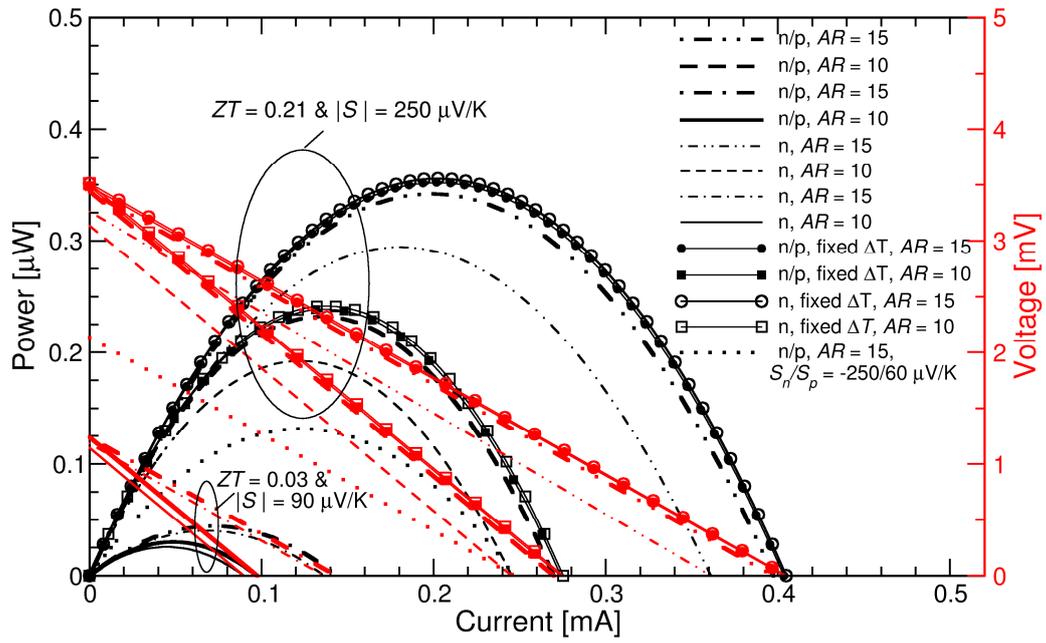

Fig. 4(b):

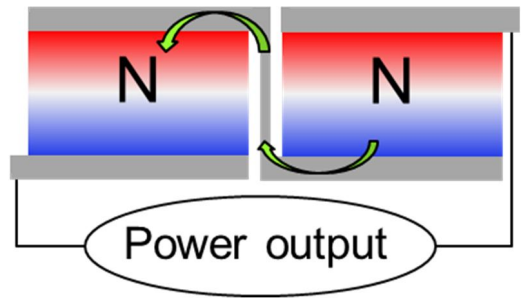

Fig. 4(c):

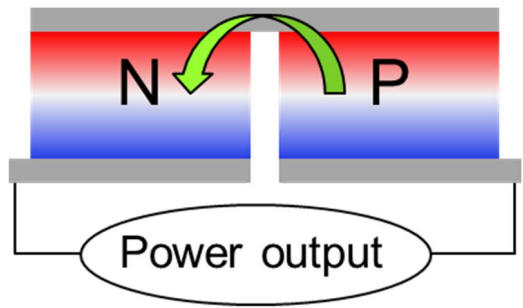

Fig. 5 (a):

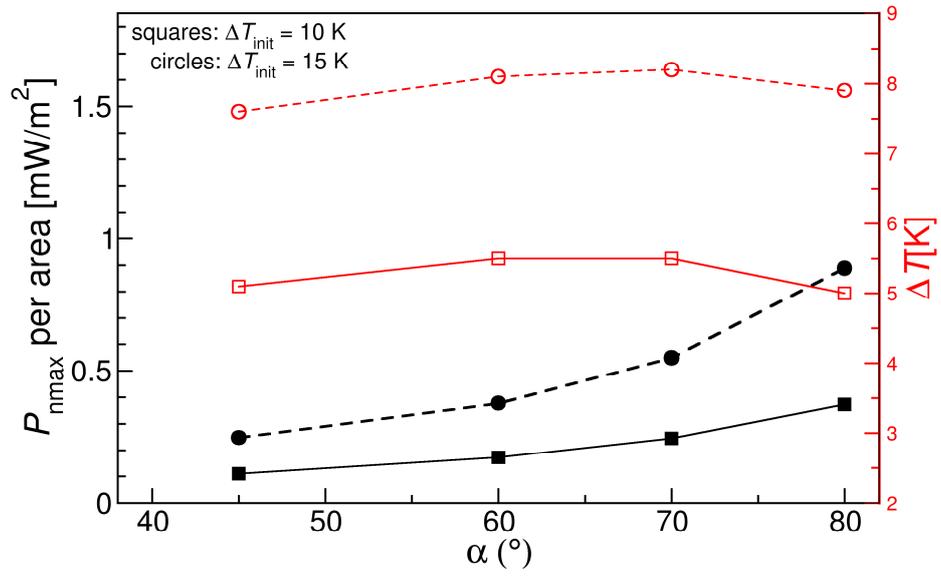

Fig. 5(b):

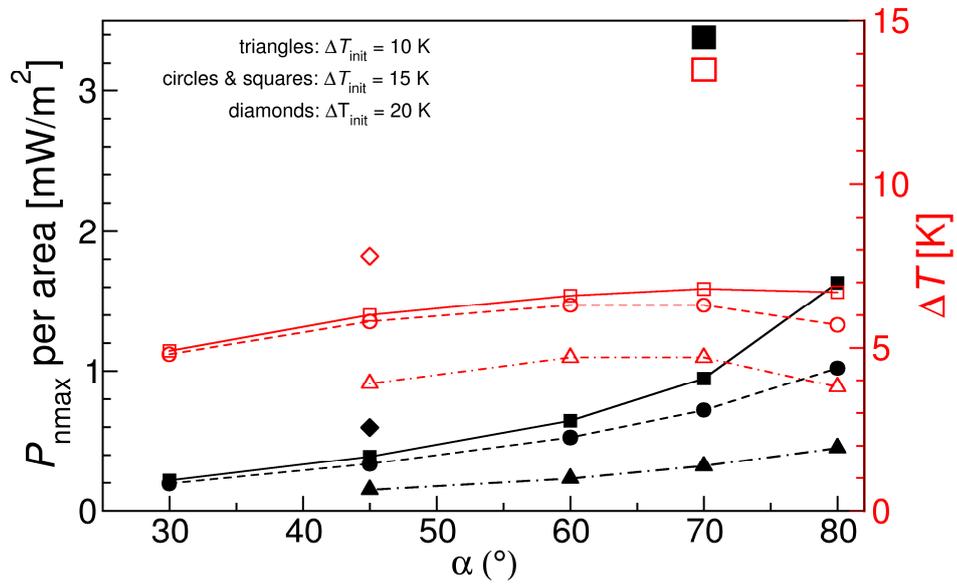

Fig. 5 (c):

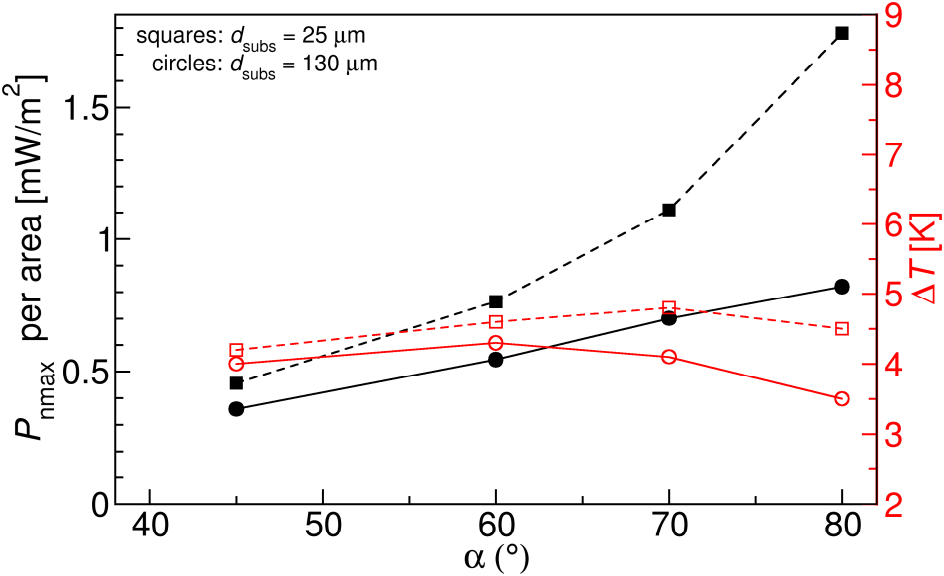